\documentclass[12pt,twoside]{article}
\usepackage[mathscr]{eucal}
\usepackage{amsmath,amsfonts,amssymb,amsthm,mathabx,empheq}
\usepackage{times}
\usepackage{pdfsync}
\usepackage{cite}
\usepackage{url}
\usepackage{hyperref}
\usepackage{tensor}
\usepackage{color}
\usepackage{multicol}
\usepackage{bbold}

\advance\textheight\topskip
\allowdisplaybreaks[1]

\newcommand\RR{\ensuremath{\mathbb R}}
\newcommand\td{\text{d}}
\newcommand\cO{{\cal O}}

\newcommand{\p}{\partial}

\newcommand{\be}{\begin{equation}}
\newcommand{\ee}{\end{equation}}
\newcommand{\bea}{\begin{eqnarray}}
\newcommand{\eea}{\end{eqnarray}}
\def\nn{\nonumber}
\def\bz{\bar z}
\def\bw{\bar w}

\def\ga{\gamma_{z\bz}}

\def\cH{{\cal H}}

\def \gai {\gamma_{z\bz}^{-1}}
\def \gawi {\gamma_{w\bw}^{-1}}

\def\n{\nabla}

\newcommand*\xbar[1]{%
  \hbox{%
    \vbox{%
      \hrule height 0.5pt 
      \kern0.3ex
      \hbox{%
        \kern-0.0em
        \ensuremath{#1}%
        \kern-0.0em
      }%
    }%
  }%
}

\DeclareFontFamily{OT1}{rsfs}{} \DeclareFontShape{OT1}{rsfs}{m}{n}{
<-7> rsfs5 <7-10> rsfs7 <10-> rsfs10}{}
\DeclareMathAlphabet{\mycal}{OT1}{rsfs}{m}{n}

\begin{document}
\title{Soft theorems in curved spacetime}

\author{Peng Cheng and Pujian Mao}

\date{}

\def\mytitle{Soft theorems in curved spacetime}

\addtolength{\headsep}{4pt}

\begin{centering}

  \vspace{1cm}

  \textbf{\Large{\mytitle}}

  \vspace{1.5cm}

 {\large Peng Cheng$^{a,b}$ and Pujian Mao$^{a}$}

\vspace{.5cm}

\vspace{.5cm}
\begin{minipage}{.9\textwidth}\small \it  \begin{center}
    ${}^a$ Center for Joint Quantum Studies and Department of Physics,\\
     School of Science, Tianjin University, 135 Yaguan Road, Tianjin 300350, China
 \end{center}
\end{minipage}

\vspace{0.2cm}
\begin{minipage}{.9\textwidth}\small \it  \begin{center}
    ${}^b$ Lanzhou Center for Theoretical Physics,\\
    Key Laboratory of Theoretical Physics of Gansu Province,\\
    Lanzhou University, 222 South Tianshui Road, Lanzhou 730000, Gansu, China
 \end{center}
\end{minipage}

\end{centering}


\vspace{1cm}

\begin{center}
\begin{minipage}{.9\textwidth}
  \textsc{Abstract}. In this paper, we derive a soft photon theorem in the near horizon region of the Schwarzschild black hole from the Ward identity of the near horizon large gauge transformation. The flat spacetime soft photon theorem can be recovered as a limiting case of the curved spacetime. The soft photons on the horizon are indeed soft electric hairs. This accomplishes the triangle equivalence on the black hole horizon.
 \end{minipage}
\end{center}
\thispagestyle{empty}


\section{Introduction}

It has recently been shown that black holes can carry soft hairs \cite{Hawking:2016msc} which provides a
possible resolution of the black hole information paradox, see,
e.g., \cite{Strominger:2017zoo} for a review and \cite{Pasterski:2020xvn,Cheng:2020vzw} for more recent developments. The discovery of the black hole soft hair is originally inspired by a fascinating equivalence relation in the infrared sector of gravity, where the BMS supertranslation Ward identity is equivalent to Weinberg's soft graviton theorem \cite{Strominger:2013jfa,He:2014laa}. In the language of S-matrix, an alternative interpretation of such equivalence relation is that the soft graviton is generated by the action of a supertranslation on the quantum state. If one implements the supertranslation action on an asymptotically flat black hole, then soft gravitons can be created on the horizon. This is the original proposal for soft hairs on black holes \cite{Hawking:2016msc}.

Though supertranslations are well defined in any asymptotically flat spacetime, the soft theorem is highly relying on a flat spacetime background, where the scattering amplitudes are defined. In curved spacetime, for instance the near horizon region of a stationary black hole, it is completely not clear if a soft graviton theorem exists. Then it is questionable if a supertranslation action can really create soft gravitons on black hole horizons. The aim of the present work is to explore soft theorems in curved spacetime from the near horizon perspective and to address their connections to soft hairs.

In this paper, we derive a soft photon theorem from the Ward identity of the near horizon large gauge transformation. In \cite{Strominger:2013lka,He:2014cra}, it was argued that the asymptotic symmetries of gauge theories are symmetries of the quantum S-matrix. We assume that this statement should be valid for any boundary of the spacetime. Considering the horizon as the inner boundary of the Schwarzschild black hole, the near horizon symmetries should be symmetries of S-matrix in the curved spacetime, at least in the near horizon region. We extend the connection of soft photon theorems and asymptotic symmetries in \cite{He:2014cra} to the near horizon region of the Schwarzschild black hole with two crucial adaptations. The first one is that we need to deal with the dispersion relation of the wave vector in curved spacetime in the mode expansion of the gauge fields. The second one is that the stationary phase approximation at null infinity (see, e.g., \cite{Strominger:2017zoo} for details) can not be applied in the near horizon region. We use the
Feynman prescription to regularize the divergence from the near horizon limit. We first write the mode expansion in the isotropic coordinates. Then the integration of the momentum in the mode expansion can be performed by the usual angular variables. Applying the Feynman prescription for regularization, we obtain the action of the charge associated to the near horizon large gauge transformation on the state without using the stationary phase approximation. Hence the soft photon theorem in position space can be derived from the Ward identity of the near horizon large gauge transformation. We transform the soft factor into momentum space and discuss its relation to the one in flat spacetime. We further show that the near horizon soft photons have precisely the forms of soft electric hairs. This is a direct proof of the statement in \cite{Hawking:2016msc} that soft particles implant soft hairs on black hole horizons. Moreover, the near horizon soft photon theorem, if it is verified from amplitudes side in curved spacetime, would accomplish the triangle relation at the horizon.

\section{Near horizon Maxwell theory}
\label{solution}

We consider Maxwell theory coupled to a conserved matter current in the near horizon region of the Schwarzschild black hole. We adopt the retarded coordinates $(u,r,z,\bz)$, where $(z,\bz)$ are the complex stereographic coordinates. The line element of the Schwarzschild black hole is
\begin{equation}
\label{metric}
\td s^2=-f(r)\td u^2-2\td u\td r+2\Omega^2\gamma_{z\bz}\td z\td\bz \ ,
\end{equation}
where
\begin{equation}
\gamma_{z\bz}=\frac{2}{(1+z\bz)^2}\ ,\quad f(r)=\frac{r}{r+2M},\quad \Omega=(r+2M).\nn
\end{equation}
The horizon $\cH$ is precisely the submanifold $r=0$ in the retarded spherical coordinates, with topology $S^2\times\RR$.

\begin{figure}[ht] 
\center{\includegraphics[width=1\linewidth]{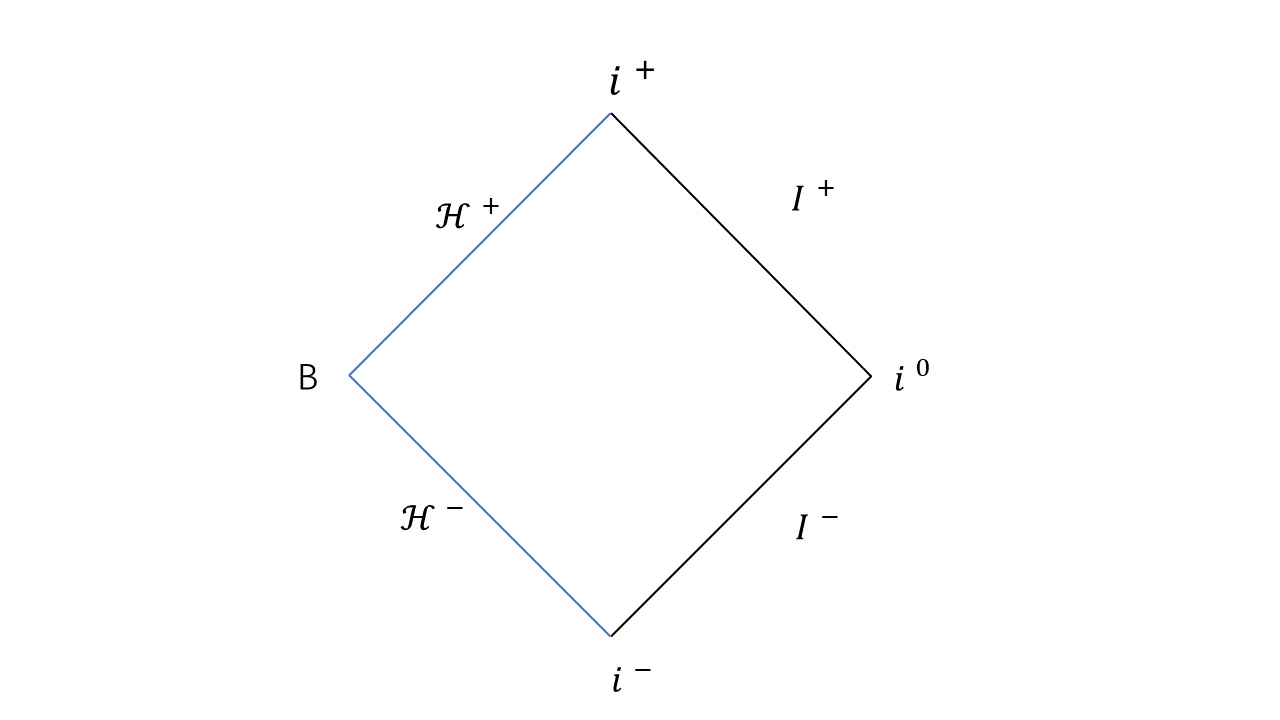}}
\caption{The Penrose diagram of the outside region of the Schwarzschild black hole.} \label{f1}
\end{figure}

The retarded coordinates can only cover half of the horizon, i.e., $\cH^-$ as demonstrated in Figure \ref{f1}. The other half $\cH^+$ can be described by advanced coordinates $(v,r,z,\bz)$. The crossing point of $\cH^-$ and $\cH^+$ is the bifurcation point $B$. In this work, we will only pay attention to the $\cH^-$ part, while everything can be
similarly repeated on $\cH^+$. The near horizon symmetries on $\cH^-$ and $\cH^+$ can be connected by the near bifurcation symmetries \cite{Adami:2020amw}.

We consider a gauge field $A_{\mu}$ and a conserved matter current $J_{\mu}$ with the radial gauge and the near horizon fall-off conditions as follows
\begin{equation}\label{asycond}
\begin{split}
&A_r=0, \quad A_u=\cO(r), \quad A_z=\cO(1), \\
&J_r=0, \quad J_u=\cO(1), \quad  J_z=\cO(1) .
\end{split}
\end{equation}
The radial component of the current is set to zero by the ambiguities of a conserved current.\footnote{When a conserved current is derived from a global symmetry, it is naturally defined up to the equivalence $J^\mu\thicksim J^\mu+\nabla_\nu k^{[\mu\nu]}$, so it makes more sense to consider equivalence classes of currents $[J^\mu]$ (see \cite{Barnich:2001jy,Conde:2016csj} for details).} This is a more consistent choice of working in the radial gauge. We assume the following ansatz as initial data for the near horizon expansion of the gauge fields
\begin{equation}
\label{Azz}
	  A_{z(\bz)}=A^0_{z(\bz)}(u,z,\bz) + \sum\limits_{m=1}^\infty  A^m_{z(\bz)}(u,z,\bz) r^m  \ ,
\end{equation}
and the components of the current
\begin{equation}
  	J_{z(\bz)}=J^0_{z(\bz)}(u,z,\bz) + \sum\limits_{m=1}^\infty  J^m_{z(\bz)}(u,z,\bz) r^{m}  \ .
\end{equation}
From the Maxwell's equations, $\nabla_\mu F^{\mu \nu}=J^{\nu}$, and the current conservation condition, $\nabla_\mu J^\mu=0$, in an $r$-expansion, we obtain
\begin{align}
\label{Ju}  &J_u=\frac{J^0_u(u,z,\bz)}{
  \Omega^2} + \frac{1}{\Omega^2} \int_0^r \td r'\left[ \gai(\p_z J_{\bz} + \p_{\bz} J_z)\right] \ ,\\
\label{Au}
  &A_u=\frac{A^0_u(u,z,\bz)}{2M} - \frac{A^0_u(u,z,\bz)}{\Omega} \nn \\
  &+ \int_0^{r} \td r'\, \frac{1}{\Omega^2} \int_0^{r'}\td r''\left[ \gai(\p_z \p_{r''} A_{\bz} + \p_{\bz}\p_{r''} A_z)\right] \ ,\\
\label{supeq}
  &\p_u A^0_u  =-\gai \p_u(\p_z A^0_{\bz} + \p_{\bz} A^0_z) - J^0_u \ ,\\
  &2\p_u\p_r A_z= \p_z \p_r A_u + \p_r (f\p_r A_z) + \p_z \left[\frac{1}{\Omega\ga} (\p_{\bz} A_z - \p_z A_{\bz})\right] - J_z\ ,\end{align}
where $J^0_u(u,z,\bz)$ and $A^0_u(u,z,\bz)$ are integration constants. We have omitted one equation that can be obtained from the last one above by simply swapping $z\rightleftharpoons\bz$. The time evolution of the coefficients in the expansion of $A_{z(\bz)}$ are uniquely determined except the leading $A^0_{z(\bz)}$.
We refer to $\partial_u A^0_{z(\bz)}$ as the electromagnetic ``\textit{news}'' since they reflect the propagation of electromagnetic waves.

The gauge and fall-off conditions in~\eqref{asycond} leave residual gauge transformations of the form $\delta A_{\mu}=\partial_{\mu}\varepsilon(z,\bz)$, where $\varepsilon(z,\bz)$ is an arbitrary function on the horizon. We define the associated charge at the bifurcation point on the horizon \cite{Barnich:2001jy},
\begin{equation}\begin{split}
Q_{\varepsilon}&= \int_{B} \td z \td\bz \,\ga\, \varepsilon(z,\bz)\, \Omega^2\,F_{ru}\\
&= \int_{\cH^-} \td z \td\bz \td u\,\ga\, \varepsilon(z,\bz)\, \Omega^2\,\p_u F_{ru},
\end{split}\end{equation}
where in the second equality we assumed $A^0_u|_{i^-}=0$, meaning that in the far past the system contains no bulk electric charge. Inserting the solution \eqref{supeq} into the charge yields
\begin{equation}\label{charge}
  Q_{\varepsilon}=-\int_{\cH^-} \td z \td\bz \td u\, \varepsilon(z,\bz)\left(\ga J_u^0 + \p_u\p_z A_{\bz}^0 + \p_u\p_{\bz} A_z^0\right) \ .
\end{equation}

\section{Near horizon soft photon theorem}

If a symmetry is generated by a charge $Q$, then the associated Ward identity in the S-Matrix language reads as
\begin{equation}
\label{Ward}
	\langle\rm{out}|Q^{\textrm{out}}-Q^{\textrm{in}}|\rm{in}\rangle=0 \ .
\end{equation}
For a spontaneously broken symmetry, the charge must act non-linearly on the states. Otherwise it would annihilate the vacuum. Hence one can decompose the charge into linear and non-linear pieces $Q=Q_{\rm{L}}+Q_{\rm{NL}}$. The Ward identity for a broken charge can be written as
\begin{equation}
\label{Wards}
	\langle\rm{out}|Q_{\rm{NL}}^{\textrm{out}}-Q^{\textrm{in}}_{\rm{NL}}|\rm{in}\rangle=
	-\langle\rm{out}|Q^{\textrm{out}}_{\rm{L}}-Q^{\textrm{in}}_{\rm{L}}|\rm{in}\rangle\,.
\end{equation}

The analysis in the previous section is carried out only near $\cH^-$. We assume that all \textit{in} states in the S-matrix are defined on $\cH^-$. Particles are created on $\cH^-$ then propagated to the bulk and finally ended on $\cH^+$. In this sense, $\cH^-$ and $\cH^+$ are very similar to the past and future null infinity $I^-$ and $I^+$. We will only paid attention to the \textit{in} part in the present work. The analysis of the \textit{out} part can be carried out analogously up to an identification at the bifurcation point. For notational brevity, we suppress the \textit{in} label in the following text.

Let us now decompose the charge~\eqref{charge} into a piece containing the ``news'' and the rest containing the sources\footnote{We drop the total minus sign in the charge as it won't affect the Ward identity.}
\begin{align}
\label{Q0NLm}
	Q_{\textrm{NL}}=\int_{\cH^-}\td z\td\bz \td u\,\varepsilon\,\partial_u\partial_{\bz}A^0_z \ ,\\
\label{Q0Lm}
	Q_{\textrm{L}}=\frac12\int_{\cH^-}\td z\td\bz \td u\,\gamma_{z\bz}\,\varepsilon\,J_u^0 \ .
\end{align}
These are, respectively, the non-linear and linear pieces above. We keep only the anti-holomorphic terms for the ``news'' part. The sources are split into two parts. That is where the half factor comes in \eqref{Q0Lm}. Such split is consistent with the treatment in~\cite{He:2014cra} where the authors consider only the sector of the phase space with no long-range magnetic fields. In our case, such conditions lead to $F_{z\bz}=0$ at $B$ and $i^0$. Correspondingly, the helicity of soft photon is decoupled. Hence one can only deal with either the holomorphic part or the anti-holomorphic part.

The soft photon theorem is obtained from the Ward identity by a concrete choice of $\varepsilon(z,\bz)$ at null infinity~\cite{He:2014cra}. We follow the same choice for the near horizon case,
\begin{equation}
\label{eps}
	\varepsilon(z,\bz)=\frac{1}{w-z} \ ,
\end{equation}
for which $\partial_{\bz}\varepsilon=-2\pi\delta^2(z-w)\,$.

\subsection{Mode expansion in Schwarzschild spacetime}

The mode expansion of fields in curved spacetime needs to be adapted to the dispersion relation of the momentum of a moving particle in curved spacetime. It is more convenient to work in the isotropic coordinates system $(t,x_i)$ where the line element of the Schwarzschild black hole is
\be\label{isoflat}
\td s^2=-\left(\frac{2\rho-M}{2\rho+M}\right)^2\td t^2 +\frac{(2\rho+M)^4}{16\rho^4}d\vec x^2\,.
\ee
The isotropic coordinates are connected to the retarded coordinates in \eqref{metric} by
\be\label{transformation}
t=u+r+2M+2M\ln \frac{r}{2M} , \quad (1+\frac{M}{2\rho})^2\rho =r+ 2M,
\ee
and
\be
x^1=\rho\frac{(z+\bar z)}{1+z\bar z},\quad x^2=\rho\frac{-i(z-\bar z)}{1+z\bar z},\quad x^3=\rho\frac{1-z\bar z}{1+z\bar z} .
\ee

The Schwarzschild solution \eqref{isoflat} has one timelike Killing vector $\frac{\p}{\p t}$.
One can define the corresponding conserved energy as
$\omega=-p_0$.
Thus, the dispersion relation for the massless particle moving along a null trajectory is
\be\label{dispersion}
-\left(\frac{2\rho+M}{2\rho-M}\right)^2 \omega^2+\frac{16\rho^4}{(2\rho+M)^4}{\vec p~}^2=0,
\ee
where $\vec p$ is the three momenta of the conformally flat part.

For a free massless scalar field $\Phi(x)$, we can write the field operator in isotropic coordinates as \cite{Mukhanov:2007zz}
\begin{equation}
	\Phi(x^\mu)=-\frac{1}{(2\pi)^4}\int
	\td\omega \td^3\vec p~\tilde{\phi}(p_\mu)~e^{ip\cdot x}.\label{expansion}
\end{equation}
Inserting the dispersion relation as a delta function in the integral and working out the integration of $\omega$
\be
	-\int \td \omega~2\pi \delta \left[-\left(\frac{2\rho+M}{2\rho-M}\right)^2 \omega^2+\frac{16\rho^4}{(2\rho+M)^4}{\vec p~}^2\right]= \left(\frac{2\rho-M}{2\rho+M}\right)^2\frac{1}{2\omega},
\ee
the expansion (\ref{expansion}) can be further written as
\begin{eqnarray*}
	\Phi(x^\mu) &=& \frac{1}{(2\pi)^3}\left(\frac{2\rho-M}{2\rho+M}\right)^2\int \frac{\td^3\vec p}{2\omega}~\tilde{\phi}(p_\mu)e^{ip\cdot x}\Big{|}_{-p_0=\omega}\\
	&=&\frac{1}{(2\pi)^3}\left(\frac{2\rho-M}{2\rho+M}\right)^2\int ~\frac{\td^3\vec p}{2\omega}~\left[\mathfrak{a}(p_\mu)e^{ip\cdot x}+\mathfrak{a}^{\dagger}(p_\mu)e^{-ip\cdot x}\right]\Big{|}_{\omega>0}.
\end{eqnarray*}
In the flat spacetime case one only has $\frac{1}{2\omega}$ in the mode expansion. Now we have an extra factor $\frac{({2\rho-M})^2}{({2\rho+M})^2}$ because of the dispersion relation \eqref{dispersion}.

For a $U(1)$ gauge field $A_{\mu}$, the mode expansion will involve two polarization vectors orthogonal to the propagating direction. The mode expansion of $A_{\mu}$, with respect to the dispersion relation \eqref{dispersion}, is
\begin{equation}
A_{\mu}(x)=\sum_{\alpha=\pm}\frac{1}{(2\pi)^3}\left(\frac{2\rho-M}{2\rho+M}\right)^2 \int ~\frac{\td^3\vec p}{2\omega}~\left[\epsilon_\mu^{*\alpha} \mathfrak{a}_{\alpha}(p)e^{ip\cdot x} + \epsilon_\mu^{\alpha}\mathfrak{a}_{\alpha}^{\dagger}(p)e^{-ip\cdot x}\right].
\end{equation}
The polarization vectors satisfy the normalization condition $\epsilon_{\alpha}^{\mu}\epsilon^*_{\beta\mu}=\delta_{\alpha\beta}$. It is of convenience to
parametrize the photon four-momenta by
\be\label{softp}
p_{\mu}= \frac{\omega}{1+z\bar z}\frac{(2\rho+M)^3}{4 \rho^2 (2\rho-M)}\left(\frac{4 \rho^2 (M-2\rho)}{(2\rho+M)^3}(1+z\bar z), (z+\bar z),-i(z-\bar z),(1-z\bar z)\right).
\ee
The polarization vector can be written as
\be\label{polarization}
\begin{split}
\epsilon^{+\mu}=& \frac{1}{\sqrt{2}}\frac{4\rho^2}{(2\rho+M)^2}\left(\frac{(2\rho+M)^3}{4\rho^2(2\rho-M)}\bar{z},1,-i,-\bar z\right),\\
\epsilon^{-\mu}=& \frac{1}{\sqrt{2}}\frac{4\rho^2}{(2\rho+M)^2}\left(\frac{(2\rho+M)^3}{4\rho^2(2\rho-M)}z,1,i,-z\right).
\end{split}\ee
In the $(z,\bar z)$ coordinates, we have
\be
\epsilon^{+}_{z}=\frac{(M+2\rho)^2}{2\sqrt{2}\rho(1+z\bar z)},\quad \epsilon^{-}_{\bar z}=\frac{(M+2\rho)^2}{2\sqrt{2}\rho(1+z\bar z)}.
\ee
Inserting those expressions and using the fact that the spatial part of the metric is conformally flat, the mode expansion can be rewritten as
\be
\begin{split}
A_{\mu} =& \sum_{\alpha=\pm}
\frac{(2\rho+M)^7}{(2\rho)^6(2\rho-M)}\int \frac{\omega~\td \omega}{8\pi^2}\int_0^{\pi}\td \Theta \\
&\times \left(\sin \Theta  \epsilon^{\alpha}_{\mu}\mathfrak{a}_\alpha e^{-i\omega t+i\omega \frac{(2\rho+M)^3}{4\rho(2\rho-M)}\cos \Theta}+c.c.\right),
\end{split}
\ee
where $\Theta$ is the angle between $\vec p$ and $\vec x$. The mode expansion in the flat spacetime (see, e.g., in \cite{He:2014cra,Lysov:2014csa,Strominger:2017zoo}) can be recovered by setting $M=0$. Near the horizon, we have $\rho\to \frac{M}{2}$. Let $R=\rho - \frac{M}{2}$. Hence $R=r+O(r^2)$ in the near horizon region. The $\Theta$ integration in the mode expansion can be worked out directly with the help of the relation
\be
\int_0^{\pi}\td \Theta \sin \Theta ~e^{i\omega \frac{2 (M+R)^3}{R (M+2 R)}\cos \Theta}=
 \frac{R (M+2 R) }{\omega  (M+R)^3}\sin \left[\frac{2 \omega  (M+R)^3}{R (M+2 R)}\right] .
\ee
The exponential in the integrand blows up as $R\to0$. At null infinity, one can apply a stationary-phase approximation in this step to control the divergence when $r\to\infty$, see, e.g., in \cite{He:2014cra,Lysov:2014csa,Strominger:2017zoo}. Here we follow the Feynman prescription to introduce a small imaginary part to the radial coordinate $R\to R+i\tau$. This will keep the radial coordinate $R$ continuous on the real axis. Then we will trace all the contributions without any approximation. For notational brevity, we will momentarily suppress the imaginary part $i\tau$. In terms of $R$, we obtain the mode expansion as
\begin{multline}
A_{\mu}
= \sum_{\alpha=\pm}
\frac{8(M+R)^4}{\pi^2(M+2R)^5} \int \td \omega ~ \left(\frac{R^2}{M^2+2 M R}\right)^{-2 i M \omega }\\
 \times \epsilon^{\alpha}_{\mu}\mathfrak{a}_\alpha e^{-i\omega u-i\omega \frac{2 (M+R)^2}{M+2 R}}\sin \left[\frac{2 \omega  (M+R)^3}{R (M+2 R)}\right]
+c.c. .
\end{multline}
Note that we have adopted the retarded time $u$ using the transformation in \eqref{transformation}.
The near horizon field $A_z^0$ is related to the plane wave modes by
\begin{multline}
A_z^0=\frac{16\sqrt{2}}{\pi^2(1+ z \bar z)}\frac{(M+R)^6}{(M+2R)^6} \int \td \omega~ \left(\frac{R^2}{M^2+2 M R}\right)^{-2 i M \omega } \\
\times   \sin \left[\frac{2 \omega  (M+R)^3}{R (M+2 R)}\right]
\left( \mathfrak{a}_{+} e^{-i\omega u-i\omega \frac{2 (M+R)^2}{M+2 R}}
+\mathfrak{a}^{\dagger}_{-} e^{i\omega u+i\omega \frac{2 (M+R)^2}{M+2 R}} \right).\label{modefinal}
\end{multline}

\subsection{Soft photon theorem in position space}

The non-linear part of the conserved charge in \eqref{Q0NLm} can be expressed as
\be
Q_{\textrm{NL}}=\lim_{R\to 0} \int_{\cH^-}\td z\td\bz \td u\,\varepsilon(z,\bz)\,\partial_u \partial_{\bz}A_{z} \,.
\ee
Inserting $\varepsilon(z,\bz)=\frac{1}{w-z}$ and using the Fourier relation
\begin{equation}
	\int_{-\infty}^{\infty}\td u\,\partial_uF(u)=2\pi i\lim_{\omega\to0}\left[\omega\tilde{F}(\omega)\right],
\end{equation}
where we define  $F(u)=\int_{-\infty}^{\infty}\td\omega\,e^{i\omega u}\tilde{F}(\omega)$, the non-linear part of charge with the mode expansion \eqref{modefinal} is reduced to
\bea
Q_{\textrm{NL}}
&=& \frac{-128\sqrt{2}i}{(1+ w \bw)} \frac{M^2}{R} \lim_{\omega\to 0}
~ \left[ \omega^2\mathfrak{a}_{+}
+\omega^2\mathfrak{a}^{\dagger}_{-} \right].
\eea
Hence, the action of this piece of the charge on the \textit{in} state is
\bea\label{QNL}
\langle\text{out}|Q_{\rm{NL}}|\text{in}\rangle
= \frac{-128\sqrt{2}i}{(1+ w \bw)} \frac{M^2}{R} \lim_{\omega\to 0}
~ \langle\text{out}| \omega^2\mathfrak{a}^{\dagger}_{-} |\text{in}\rangle
\eea

Regarding the linear piece \eqref{Q0Lm},
for simplicity, we restrict ourselves to
scalar charged (with charge $Q_e$) matter with current $J_\mu=i Q_e(\bar{\Phi}\n_\mu \Phi - \Phi\n_\mu \bar{\Phi})$. Since the derivatives act on complex scalar field $\Phi$, it makes no difference whether using covariant derivative or normal derivative. With respect to \eqref{Ju}, the current at leading order is $J^0_u=i \Omega^2 Q_e(\bar{\Phi}\p_u \Phi - \Phi\p_u \bar{\Phi})$.
Then one can apply the canonical commutation relation similar to the flat spacetime case \cite{Lysov:2014csa}
\begin{equation}
  [\bar{\Phi}^0(u,z,\bz),\Phi^0(u',w,\bw)]=\frac{i}{4}\Omega^{-2}\gawi\, \Theta(u-u')\delta^2(z-w) \,,
\end{equation}
to get the action of the linear charge on the \textit{in} state as
\be
\label{QmL}
  \langle\textrm{out}|Q^{(0)}_{\textrm{L}}|\textrm{in}\rangle=\sum_{k=1}^n
  \frac{ Q_k}{4(w-w_k)}\langle\textrm{out}|\textrm{in}\rangle,
\ee
where $Q_k$ is the electric charge of the $k$-th particle. We have assumed there are $n$ incoming particles.

Finally the Ward identity \eqref{Wards} will yield a soft photon theorem in Schwarzschild spacetime as
\be\label{presoftT}
\lim_{\omega\to 0}
~ \langle\text{out}| \mathfrak{a}^{\dagger}_{-}|\text{in}\rangle =
\frac{1}{512}\frac{R}{i\omega M^2}\left(\frac{1+ w \bw}{\sqrt{2}\omega}
\sum_{k=1}^n\frac{Q_k}{w-w_k}\right)
\langle\textrm{out}|\textrm{in}\rangle.
\ee
We have several remarks on this soft theorem as follows:
\begin{itemize}
\item
A priori, one can not deduce that the Ward identity of the near horizon symmetry must yield a soft theorem, though it does reveal some relations from the S-Matrix. The derivation of the soft theorem is particularly based on the form of the charge \eqref{charge} from near horizon analysis.

\item
We use the Feynman prescription to regularize the radial coordinate $R$ on the horizon. A similar treatment is not needed at null infinity because the null infinity is conformally introduced. The divergence from $r\to\infty$ can be absorbed into the conformal factor. Hence the quantities defined at the null infinity should be finite, but the horizon is part of the spacetime, and particles can freely cross it. So the divergence from the horizon should be regularized. On the horizon, we set $R=i\tau$ and $\tau$ being infinitesimal.

\item
Let $k=\frac{\tau}{512\omega M^2}$ be a finite constant. The soft factor in \eqref{presoftT} recovers the flat spacetime soft photon factor when $k=1$.

\end{itemize}

\subsection{Soft photon theorem in momentum space}

We consider the \textit{in} states (the hard particles) are created on the horizon. Then the hard momenta in the near horizon region can be parametrized as
\be
q_{k\mu} =\frac{E_k}{1+z_k\bar z_k}\frac{4M}{R}
\left(-\frac{R}{4M}(1+z_k\bar z_k), (z_k+\bar z_k),-i(z_k-\bar z_k),(1-z_k\bar z_k)\right).
\ee
Using the soft momentum in \eqref{softp} and the polarization vectors in \eqref{polarization}, one can show that
\be\label{useful}
\sum_{k=1}^{n} \frac{Q_k q_k\cdot \epsilon}{q_k\cdot p}=\frac{R}{\omega M}
\left(\frac{1+z\bz}{\sqrt{2}}\sum_{k=1}^{n} \frac{Q_k }{z-z_k}\right).
\ee
Applying this relation, we obtain the soft photon theorem in momentum space as
\be\label{softmomentum}
\lim_{\omega\to 0}
~ \langle\text{out}| \mathfrak{a}^{\dagger}_{-}|\text{in}\rangle=
\frac{1}{512}\frac{1}{i\omega M} \sum_{k=1}^{n} \frac{Q_k q_k\cdot \epsilon}{q_k\cdot p}\langle\textrm{out}|\textrm{in}\rangle .
\ee
To close this section, we will comment on two limits of the black hole mass parameter $M$. One can read from \eqref{softmomentum} that the soft factor is singular when $M=0$. Though setting $M=0$ will not bring any divergence in the computation, it will lead to the fact that $R=0$ is just a point rather than a three dimensional null hypersurface. There is no horizon at all at $R=0$ in such case. Somehow the regularization in $R\to R+i\tau$ when $R=0$ prevents one taking the limit $M\to0$. Consequently, we need to introduce a corresponding regularization $M\to M+im$, where $m$ is infinitesimal. Another limit is to set $M\to\infty$. Of course, this case may not be physically sound, it can provide a mathematical consistency check for the soft theorem. In this limit, one can arrange that $512\omega M=1$. Hence the flat space soft theorem is recovered. Naively, it is a curious fact that the flat space result is not recovered from the flat space limit $(M=0)$. Nevertheless, the radius of the horizon is proportional to the mass $M$. Technically, a large mass limit requires a large radius to compensate. This makes the near horizon analysis very close to the null infinity analysis. Since the flat space soft theorem can be derived from asymptotic symmetry at null infinity, it is reasonable to see the flat space result is from the large mass limit.

\section{Soft photons as soft electric hair on the horizon}

\subsection{Near horizon electromagnetic memories and soft electric hairs}

We follow the prescription in \cite{Adami:2021nnf} to formulate the near horizon electromagnetic memories. Consider that an electromagnetic shockwave passes through the horizon of the black hole. The information about the electromagnetic wave is encoded in the change of the surface charge. Suppose that there is initially no electromagnetic field and the system finally settles into a stationary solution with flat gauge connections on the horizon $\p_z A_{\bz}^0 = \p_{\bz} A_z^0$ \cite{He:2014cra}. According to \eqref{charge}, the change of the surface charge is just the charge evaluated at a later time,
\begin{equation}\label{changeQ}
\Delta  Q_{\varepsilon}=-\int_{u=u_f} \td z \td\bz \, \varepsilon(z,\bz)\left( \p_z  A_{\bz}^0 + \p_{\bz} A_z^0\right) \ ,
\end{equation}
where we only consider the contribution from electromagnetic shockwave. The permanent change of the surface charge \eqref{changeQ} is called the near horizon electromagnetic memory.

Alternatively, the near horizon electromagnetic memory can be derived from the near horizon large transformation as a precise example of the equivalence between memories and asymptotic symmetries \cite{Hawking:2016msc,Hawking:2016sgy,Donnay:2018ckb}. The action of the infinitesimal near horizon large transformation, denoted by $\chi(z,\bz)$, on the electromagnetic fields is
\be
\delta_\chi A_z=\p_z \chi(z,\bz),\quad \delta_\chi A_{\bz}=\p_{\bz} \chi(z,\bz).
\ee
For the case of initially no electromagnetic field, the final electromagnetic fields are just
\be
A_z^0=\p_z \chi(z,\bz),\quad  A_{\bz}^0=\p_{\bz} \chi(z,\bz).
\ee
The change of the surface charge \eqref{changeQ} is
\begin{equation}
\Delta  Q_{\varepsilon}=-2\int_{u=u_f} \td z \td\bz \, \varepsilon(z,\bz) \p_z\p_{\bz} \chi .
\end{equation}
Since large gauge transformations create soft hairs on the black hole horizon \cite{Hawking:2016msc,Hawking:2016sgy,Donnay:2018ckb}, the electromagnetic shockwave can indeed implant soft electric hairs on the black hole horizon. As we will show in the next subsection, the soft electric hairs can be implanted by soft photons on the horizon.

\subsection{Near horizon soft photons as soft hairs}

Soft photon theorems connect an $n+1$ particles state to an $n$ particles state in a low-energy expansion of the extra photon in the form
\be
\label{soft}
M_{n+1}\big(p_1,\ldots,p_n,\left[q;\epsilon^{\pm}\right]\big)=
\frac1\omega S^{(0)\pm} M_{n}(p_1,\ldots,p_n)+\cO\left(1\right).
\ee
In the low-energy limit, the soft theorem should have certain relevance to classical computation. Following the proposal in  \cite{Strominger:2014pwa,Mao:2017wvx} at null infinity, one can read from the soft theorem that the soft factors can be interpreted as the expectation value of near horizon fields fluctuation,
\be
S^{(0)\pm}=\lim_{\omega\rightarrow0} \omega\frac{M_{n+1}^{\pm}}{M_{n}}.
\ee
Then the expectation value of near horizon fields fluctuation with soft photon emission is simply related to the change of classical fields as
\be\label{leading}
A_z=\epsilon^{*+}_z S^{(0)+} + \epsilon^{\ast-}_z S^{(0)-},
\ee
where we assume the system initially has no electromagnetic field.

Inserting the soft photon factors \eqref{presoftT} into \eqref{leading} (recall the complex conjugate), we obtain
\be
A_z=2k\sum_{k=1}^{n}\frac{Q_k}{z-z_k},\quad
A_{\bz}=2k\sum_{k=1}^{n}\frac{Q_k}{\bz-\bz_k}.
\ee
Consider $k=\frac{\tau}{512\omega M^2}$ as a constant factor, the above electromagnetic fields are exactly the form of a memory discussed previously in this section. Hence the soft photons in the near horizon region naturally have the form of soft electric hairs.
Inserting the gauge fields into the charge \eqref{changeQ}, one obtains
\begin{equation}
\Delta  Q_{\varepsilon}=-8\pi k\sum_{k=1}^{n} Q_k \varepsilon(z_k,\bz_k),
\end{equation}
where $(z_k,\bz_k)$ are the locations of the hard particles projected on the horizon, which generate the electromagnetic shockwave.

\section{Conclusion and discussion}

We derive a soft photon theorem in the near horizon region of the Schwarzschild black hole. The soft factor can recover the flat spacetime soft photon factor in the zero mass limit $M\to\infty$. The soft factor interpreted as the expectation value of near horizon
fields fluctuation is in the form of soft electric hairs on the black hole horizon. Hence this verifies that the soft particles on black hole horizons can implant soft hairs.

In the derivation of the soft photon theorem, we restrict ourselves to the case that all particles are in the near horizon region, but we expect our analysis to be generic to the formalism and relevant issues can be addressed elsewhere. We point out some for future directions. The most direct one is to extend our derivation to other curved spacetime with horizon, such as a Kerr black hole or cosmological solution. Another glaring one is to study the scattering with particles in both a near horizon region and asymptotic region. In particular, one may need to enhance the near horizon or asymptotic symmetries to symplectic symmetries \cite{Compere:2015knw}. Inspired by the sub-leading soft photon theorem \cite{Low:1958sn} (see also \cite{Lysov:2014csa}) in the flat spacetime, one may wonder if there are more soft photon theorems in the near horizon region. It is more promising in the case of gravity, where one has two types of near horizon symmetries, namely near horizon supertranslations and superrotations \cite{Donnay:2015abr}. Naively, one should be able to derive two soft graviton theorems in the near horizon region. Another interesting point in the gravity case is to include higher derivative terms. They are order-suppressed in the asymptotic region, see also discussions in \cite{Elvang:2016qvq} from the soft theorem side. However in the near horizon region, the higher derivative terms have significant influence on the near horizon charges \cite{Liu:2022uox}. If they can affect the near horizon soft graviton theorem, then that would be a very meaningful investigation.

\section*{Acknowledgments}

The authors thank Laura Donnay, Zhengwen Liu, Shahin Sheikh-Jabbari, and Jun-Bao Wu for useful discussions. The authors thank Kai-Yu Zhang for pointing out the mistake in the commutation relation of the scalar fields. This work is supported in part by the National Natural Science Foundation of China (NSFC) under Grants No. 11905156 and No. 11935009. P.C. is also supported in part by NSFC Grant No. 12047501.

\bibliography{ref}

\end{document}